# Low Temperature Nanoscale Electronic Transport on the MoS$_2$ surface

R. Thamankar[1], T. L. Yap[1,2,3], K. E. J. Goh[1], C. Troadec[1] and C. Joachim[1,4]

[1] Institute of Materials Research and Engineering, Agency for Science, Technology and Research, 117602 Singapore.

[2] Department of Physics, National University Singapore, 2 Science Drive 3, Singapore 117542.

[3] GLOBAL FOUNDRIES Singapore Pte Ltd, 60 Woodlands Industrial Park, Singapore.

[4] CEMES, CNRS, 29 rue J. Marvig, 31055 Toulouse Cedex, France.

**Abstract:**

Two-probe electronic transport measurements on a Molybdenum Disulphide ($MoS_2$) surface were performed at low temperature (30K) under ultra-high vacuum conditions. Two scanning tunneling microscope tips were precisely positioned in tunneling contact to measure the surface current-voltage characteristics. The separation between the tips is controllably varied and measured using a high resolution scanning electron microscope. The $MoS_2$ surface shows a surface electronic gap ($E_S$) of 1.4eV measured at a probe separation of 50nm. Furthermore, the two- probe resistance measured outside the electronic gap shows 2D-like behavior with the two-probe separation.

Two dimensional (2D) atomic crystals have shown promising electronic properties which can be used for future mesoscopic electronic devices[1,2,3,4,5]. Monolayers of transition metal dichalcogenides ($MX_2$ where M = Metal, X = S, Se and Te) are currently in the spotlight because they show interesting 2D electronic properties and atomically clean surfaces can be obtained by the mechanical exfoliation technique. A model system in this dichalcogenide family is molybdenum disulphide ($MoS_2$). It has an indirect bulk electronic band gap $E_B \sim 1.3eV$ and an indirect – direct band gap transformation occurs when $MoS_2$ is thinned down to its monolayer limit. In this case, calculations suggest that $E_B$ increases from its bulk value of 1.3eV to 1.9eV for a single $MoS_2$ layer[6,7,8,9,10]. Various studies have been reported using few monolayers of $MoS_2$ as active material in devices, such as field effect transistors[11,12,13] sensors[14], phototransistors[15] and integrated circuits based on bilayer $MoS_2$ transistors[6].

The $MoS_2$ surface holds promise for the construction of surface atomic scale circuits. By means of a scanning tunneling microscope (STM), sulphur atoms can be extracted one by one from the $MoS_2$ surface[16]. The missing surface atom then creates a localized electronic state in the surface electronic gap[17, 18]. By systematically removing the sulphur atoms along the surface, it was calculated that long atomic wires can be formed by sulphur atom vacancies and calculations show that this will introduce a surface conduction band located in the $MoS_2$ surface electronic gap[19,20]. This approach has the advantage of working with a stable atomic wire at room temperature as compared to the recently reported dangling bond atomic wires constructed on passivated semiconductor surfaces[21]. In the prospect of using $MoS_2$ surface atomic wires to construct atomic scale surface boolean logic circuits[22] or to connect a single molecule in a planar configuration[23], we demonstrate the measurement of transport properties on $MoS_2$ surface at a nanoscale regime comparable to the surface dimension of those atomic scale electronic circuits under ultrahigh vacuum (UHV) conditions.

In this letter, electronic properties of the $MoS_2$ surface measured using STM two-probe current-voltage (I-V) spectroscopy is demonstrated. The measurements were done on a UHV-Nanoprobe (Omicron Nanotechnology) which consists of four STM tips and a high resolution scanning electron microscope (SEM)[24]. Each STM tip can perform constant current imaging and I-V spectroscopy independently. The in-situ high resolution SEM (resolution~3.4nm) is operated at a large working distance (13mm to 15mm) to allow free and independent navigation of all the STM tips on the sample surface. We use chemically etched tungsten tips with an average apex diameter ranging between 30nm to 50nm as measured by SEM. All the measurements presented below were performed in UHV and at 30K which is the lowest temperature achievable on this instrument. Ex-situ room temperature Hall effect measurements show that our $MoS_2$ sample is natively *p*-doped ($10^{16}$ atoms/cm$^3$). No further heat treatment or cleaning procedure is applied to the surface before making the measurements. In conventional transport measurements involving devices using micro (or nano) lithography techniques to fabricate the surface contact pads, contamination induced by the resists or surface modifications induced by e-beam lithography cannot be neglected[25]. In order to avoid all these issues, we measure the transport characteristics in UHV by using individual STM tips as electrodes in top contact configuration. Furthermore, by using the high resolution SEM attached to the chamber (operated at low current), we can continuously vary the tip-to-tip distance down to a few tens of nanometers (50nm in the present case). The UHV cleanliness highlighted here is a pre-requisite for successful atom-by-atom extraction to construct the atomic wires and this is not typically catered for in standard nanolithography techniques[25].

In order to measure the electronic energy gap of the bulk MoS$_2$, experiments were done with a single tip in tunneling contact mode using an individual feedback loop. This is an essential step to identify possible bulk leakage current under the top MoS$_2$ layer. For this measurement, we intentionally maintained a tunneling barrier between the tip and the MoS$_2$ surface to form a nanoscale controllable metal-insulator-semiconductor (MIS) junction at each tip location. This tunneling barrier includes the tip-to-MoS$_2$ vacuum tunneling barrier and the remaining oxide (if any) on the tip apex. The length ($L$) of this tunnel barrier can be controlled by moving the tip away or closer to the surface with a precision better than 0.1nm at 30K. Notice that our objective is not to evaluate the maximum current flowing on the MoS$_2$ surface between the two tips, but to determine the bulk energy gap ($E_B$) and the surface electronic gap ($E_S$) in the nanoscale regime.

To tune a given tip to MoS$_2$ surface tunnel contact, its I-V characteristics were recorded at various tip-sample distances ($L$) using a single tip in the normal STM mode of operation. As presented in Fig. 1a, bulk MoS$_2$ presents an electronic energy gap ($E_B$) around 1eV for tunneling current set point between 10pA and 200pA. By increasing the tunneling set point current, the energy gap decreases from $E_{B, 10pA}$ = 1.2±0.1 eV to about $E_{B, 200pA}$ = 0.9±0.1 eV. This decrease is attributed to the change in the tip sample separation by more than 0.1nm. An interesting feature of the single tip I-V curves is the appearance of a shoulder in the range of 1.3eV to 1.8eV on the positive side of each I-V curves (see Fig.1a). We attribute this to the Molybdenum (Mo) 4$d$ and Sulphur(S) 3$p$ states of the MoS$_2$ layer. Inverse photoemission spectroscopy on the MoS$_2$ surface showed similar features which are coming from the antibonding Mo 4$d$ and S 3$p$ states[26]. Theoretical studies have also indicated that Mo $d$- S $p$ hybridization dominates in the electronic properties of MoS$_2$ near the Fermi level[19, 27].

The I-V characteristics in Fig. 1a can be interpreted using a metal-insulator-semiconductor (MIS) junction model that takes into account the thermionic emission effect, the tunneling contact and the tip induced band bending. We consider all these effects via an effective Schottky barrier height ($\phi_b$) of this MIS contact[28]. For a bias voltage V between the tip and the MoS$_2$ sample, the current through this MIS junction is given by:

$$I(V, \phi_b) = AA^*T^2 \exp(-\beta L) \exp\left(-\frac{\phi_b}{kT}\right) \exp\left(\frac{qV}{nkT}\right) \times \left[1 - \exp\left(-\frac{qV}{kT}\right)\right] \quad (1)$$

where $A^* = \frac{4\pi q m^* k^2}{h^3}$ is the effective Richardson constant. Due to the chosen tunnel contact mode, an $\exp(-\beta L)$ term was introduced in (1) where $\beta$ is the tunneling inverse decay length and $L$ the effective length of the tip-MoS$_2$ surface tunnel barrier. $k$ is the Boltzmann constant, $T$ is the temperature and $n$ the ideality factor. In an MIS junction, the ideality factor indicates the deviation from ideal Schottky barrier (i.e. $n = 1$). The effective mass of the electron $m^* = 0.48 m_0$ of the bulk MoS$_2$ is considered here[27] and $A$ is the effective tunnel contact area of the tip - MoS$_2$ surface junction.

It is well documented that band bending induced by a metallic tip arises in the case of semiconductors surfaces[30]. In the present case of well controlled MIS junction, we compare the dI/dV curves for two set points: 10pA and 200pA. The dI/dV curves obtained from the numerical differentiation of the I-V characteristics are presented in Fig. 1b. Here the shoulders appearing in the I-V curves due to Mo 4*d* - S 3*p* have not been taken into account since we focus only on the onset of the bands. At a setpoint of 10pA, a linear onset of both the valence (VB) and conduction bands (CB) can be observed. Figure 1b shows that when the STM tip is brought closer to the MoS$_2$ surface by increasing this set point current to 200pA, the top of the MoS$_2$ VB edge is affected more than the CB edge by the metal tip approach. Following the analysis of Feenstra

et.al[29, 30], for a tip radius of 30nm, with a contact potential of -1.0eV and a tip-sample separation $L$ = 1nm, the valence band bending in MoS$_2$ is rather large and can extend up to the tip Fermi level. Furthermore, using (1) and according to Fig. 1a, we estimate the value of the barrier height to be $\phi_1$ = 0.51eV which is comparable to the extracted values for various metals in contact with MoS$_2$ in transistor configurations[31].

Two-tip surface I-V measurements were done by electrically isolating the back of the MoS$_2$ sample from the ground instead using the virtual ground of the two STM I-V convertors. After every two-probe I-V measurements, the sample ground was automatically reconnected before the feedback loop of each STM tip was reactivated so as to control each tip surface distance. The lateral tip-to-tip separations were determined using the in-situ UHV-SEM of the Nanoprobe as presented in Fig. 2. Here, the SEM image shows two tips with an apex diameter of ~ 50nm (see Fig. 2c). For all the two-probe I-V measurements, Tip-2 was kept at a fixed position on the surface of the MoS$_2$. Tip-1 was controllably moved keeping its feedback loop active and monitoring its position using the UHV-SEM. In this configuration, current flows from Tip-1 to Tip-2 in a floating ground mode of operation giving rise to I-V characteristics of the MoS$_2$ surface. Figure 2a shows a series of such I-V measurements using two tips separated by a distance of ~ 400nm. For this large separation, while keeping Tip-1 at 10pA feedback loop setpoint current, the surface electronic gap E$_S$ is 1.3eV as confirmed by its dI/dV curve presented in Fig. 2b. As the Tip-1 set point current was increased from 10pA to 250pA, Tip-1 approaches to the MoS$_2$ surface (inset Fig. 2a.) and this is reflected by the I-V curve changing drastically to shows a smaller surface electronic gap. The dI/dV curve at 250pA setpoint current show that the surface electronic gap is E$_S$ = 0.5eV. This reduction of the surface electronic gap is attributed to the band bending resulting from the approach of the metallic Tip-1 to the surface. As in the case of single tip measurements (Fig. 1), the surface valence band edge is affected more than the conduction band edge.

As already mentioned, each tip-MoS$_2$ junction can be modeled by a MIS junction. In the case of a tungsten tip-MoS$_2$ junction and for a doping level of $10^{16}$ atoms/cm$^3$, the surface depletion regions start to overlap for tip-to-tip separation around $d \sim 150$nm[28]. For larger distance $\underline{d}$, as presented in Fig. 2c and 2d, the lateral tip induced surface band bending regions do not overlap. For tip separations smaller than the depletion width, an almost flat band situation occurs at both the CB and VB as schematically depicted in Fig. 2f. This leads to an almost symmetric two tip I-V curves. This effect was measured with a tip-to-tip separation of 50nm as presented in Fig. 2e where an abrupt increase of the tunneling current intensity is observed in both forward and reverse bias voltages. In this case, the measured MoS$_2$ surface electronic gap is now controlled by the overlap of the two depletion regions. This leads to an apparent electronic gap E$_S$ = 1.4eV which is larger than the bulk energy gap E$_B$.

The measured two-tip I-V curves can be modeled by a linear combination of two back to back MIS contacts (Fig. 2d). Following (1), the resulting tunneling current intensity can now be written as:

$$I(V, \emptyset_{B1}, \emptyset_{B2}) = I_0^1 \exp(-V) \exp\left(-\frac{\emptyset_{B1}}{n_1}\right) \exp\left(-\frac{V}{n_1 V_T}\right) \left\{1 - \exp\left(\frac{V}{V_T}\right)\right\} + I_0^2 \exp(-V) \exp\left(-\frac{\emptyset_{B2}}{n_2}\right) \exp\left(\frac{V}{n_2 V_T}\right) \left\{1 - \exp\left(\frac{V}{V_T}\right)\right\} \quad (2)$$

Here we consider $\emptyset_{B1} \sim (\emptyset_b + \Delta 1)$ and $\emptyset_{B2} \sim (\emptyset_b + \Delta 2)$ as the effective Schottky barriers at the Tip1-MoS$_2$ and Tip2-MoS$_2$ junctions respectively where $\Delta 1$ and $\Delta 2$ are the local change in the barrier height at each junction due to the different contacts at the two tips on the surface. We consider $n_1$ and $n_2$ as the equivalent ideality factors, which contains the contribution from the non-ideal Schottky contact. This can also account for the small correction $\Delta 1$ and $\Delta 2$ for the individual effective Schottky barrier relative to the already measured $\phi_1$. As shown in Fig. 2e, the I-V characteristics can be very well fitted using (2). By fitting the ln(I) vs V curve, we obtain large values (~ 2.2) for the equivalent ideality factors ($n_1$ and $n_2$). This large value clearly shows that our tunneling contacts are far from being ideal Schottky barriers. Ideality factors greater than $n = 1.5$ at room temperature and increasing at low temperatures have been measured for MIS junctions using molecular insulators[32,33]. In this case, the departure from $n = 1$ is attributed to recombination effects at the interface[34,35]. Detailed understanding of non-ideality requires further investigation of such nanoscale MIS junctions which is beyond the scope of this Letter.

For bias voltages outside the electronic gap, the variation of the two-tip junction resistance ($R_{2P}$) is plotted in Fig. 3 as a function of the tip separation (d) up to a very large distance d=2μm. As presented in Fig. 3, $R_{2P}$ can be well fitted using a 2D-like surface resistance variation[36]:

$$R_{2P} = \frac{R_C}{\pi} \ln\left[\frac{3d-r}{r}\right] \qquad (3)$$

Here $R_C$ is the *effective contact resistance* which includes the two MIS junctions and the surface resistance of the MoS$_2$ between the two tips. $r$ is the tip apex radius of curvature and 'd' is the separation between the tips on the surface. Assuming that the Tip-1 and Tip-2 apex radii are the same in our experiments i.e. $r = 30$nm, one finds $R_C = 1.45$GΩ for the best fit using (3). At small d, the 2D character of the large voltage surface resistance demonstrates how the leakage current is located at the MoS$_2$ top surface limiting the bulk contribution due to MoS$_2$ lamellar structure in the transverse direction.

In conclusion, we have determined the surface electronic gap of MoS$_2$ surface at the nanoscale down to a separation of 50nm between the tips. At large tip separations, the MoS$_2$ surface presents a surface electronic gap $E_S = 1.3$eV very close to the bulk MoS$_2$ band gap. At a smaller tip separation (~50 nm), the measured surface electronic gap $E_S$ =1.4eV. At large tip separations, the MoS$_2$ surface gap is impacted more by the tip-induced band bending, whereas at a small tip separation of 50nm, the tip induced band bending is minimized by the significant overlap of the surface depletion regions under the two tips. For bias voltages outside the surface electronic gap, the variation of surface resistance as a function of the tip-to-tip separation is in agreement with a 2D surface conduction mechanism. Our measurements clearly demonstrate that existence of a large surface electronic energy gap on MoS$_2$ even if nanoscale surface metallic contacts were present. This opens the way to characterize the electronic properties of surface atomic scale wires and circuits constructed on the MoS$_2$ surface.

**Acknowledgements:** Authors acknowledge financial support under the A*STAR's VIP Atom Technology project 1021100972 and EU ATMOL project contract number 270028 financial support. T. L. Yap thanks National University of Singapore, GLOBAL FOUNDRIES Singapore for the financial support during Ph.D. programme. We thank M. Bossman, for critical proofreading.

Figure (1): (a) Single tip I-V curves of *p*-MoS$_2$ at various tip-sample distances (corresponding to setpoint currents of 10pA, 30pA, 50pA, 100pA, 150pA, 200pA respectively). Measurements were done at 30K and the tip bias was set at -2.1volts. (b) The dI/dV curves (numerically differentiated after fitting the I-V curve) for 10pA and 200pA. As the tip approaches the surface, closing of the bulk energy gap can be seen. Linear onset of the VB and CB are indicated by the dotted lines. All curves shown here result from of single I-V scans and no averaging was performed.

Figure (2): Surface electronic transport measurements on MoS$_2$ surface at 30K. (a) Two probe I-V curves at tip separation~400nm for various Tip1-sample distances (corresponding to setpoint currents of 10pA, 80pA, 100pA, 150pA, 200pA, 250pA (red)). Tip1 bias was set at -2.1V. Tip2 is at 23mV/500pA. (b) The differentiated I-V curves for 10pA (black) and 250pA(red). The dotted lines are guide to the eye. The two tips used for these measurements are shown in the SEM image in(c). Both tips have an apex diameter~50nm. (d) Schematic diagram of the two MIS contacts separated by distance 'd'. For large tip-to-tip separations, the depletion regions do not overlap. (e) Two tip I-V curve for a tip separation of 50nm (red). At this distance, the MoS$_2$ surface shows a surface gap of 1.4eV. This curve can be fitted (black curve) by considering equation(2). At this probe separation, the depletion regions overlap resulting in the flat bands of MoS$_2$ as shown in (f).

Figure (3): Dependence of the series resistance (R$_{2P}$) with probe separation measured using two probes. Solid circles (red and pink) show the experimentally measured R$_{2P}$. The blue curve is the fit according to equation(3) assuming both tips with r =30nm and R$_C$=1.45GΩ for the fit.

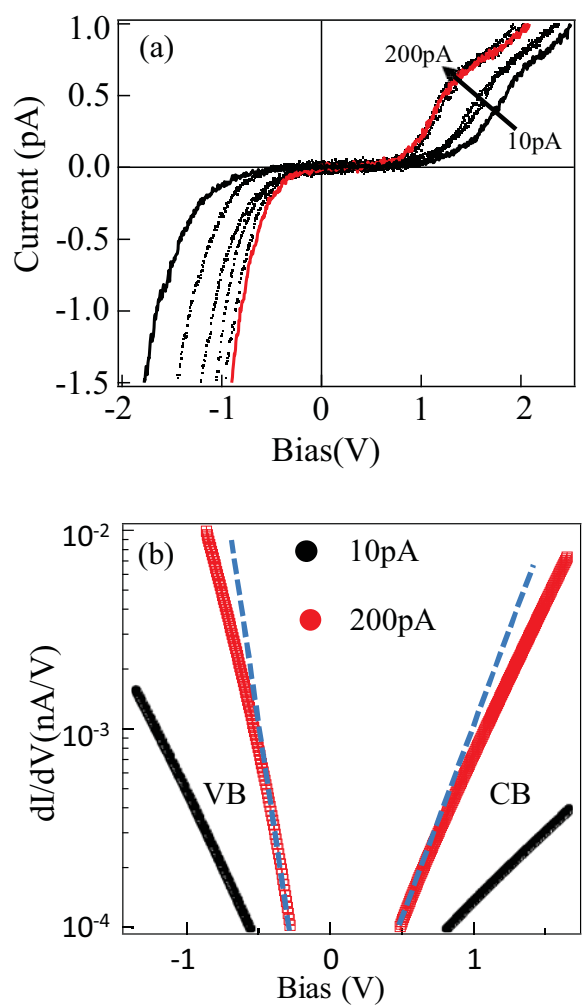

Fig (1) : (a) Single tip I-V curves of *p*-MoS$_2$ at various tip-sample distances(corresponding to setpoint currents of 10pA, 30pA, 50pA, 100pA, 150pA, 200pA respectively). Measurements were done at 30K and the tip bias was set at -2.1volts. (b) The dI/dV curves (numerically differentiated after fitting the I-V curve) for 10pA and 200pA. As the tip approaches the surfaces, closing of the bulk energy gap can be seen. Linear on set of the VB and CB are indicated by the dotted lines. All curves shown here result from of single I-V scans and no averaging was performed.

Fig (1)

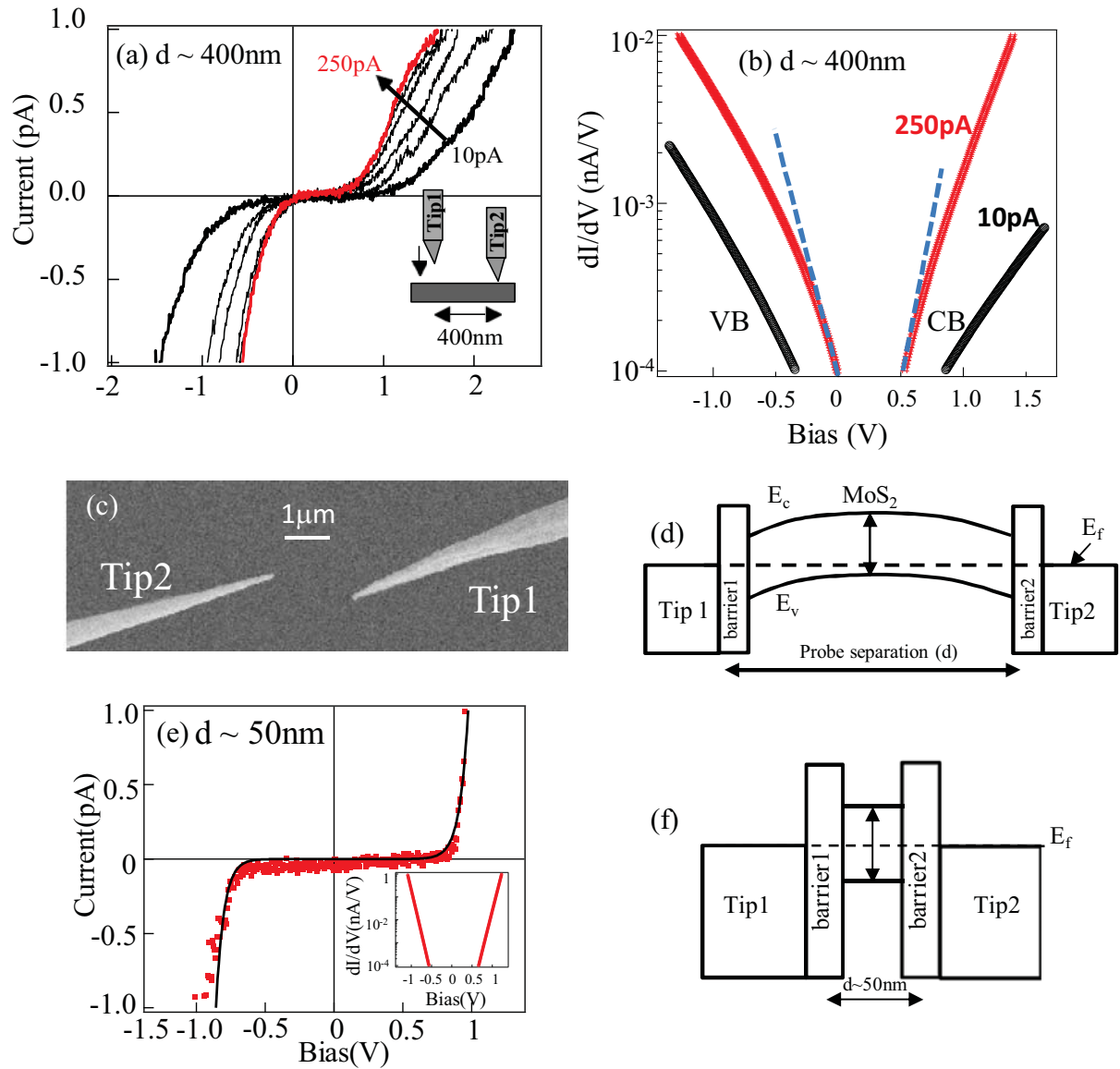

Fig (2) : Surface electronic transport measurements on MoS$_2$ surface at 30K. (a) Two probe I-V curves at tip separation ~ 400nm for various Tip1-sample distances (corresponding to setpoint currents of 10pA, 80pA, 100pA, 150pA, 200pA, 250pA(red)). Tip1 bias was set at -2.1V. Tip2 is at 23mV/500pA. (b) The differentiated I-V curves for 10pA (black) and 250pA (red). The dotted lines are guide to the eye. The two tips used for these measurements are shown in the SEM image in (c). Both tips have an apex diameter ~ 50nm. (d) Schematic diagram of the two MIS contacts separated by distance 'd'. For large tip-to-tip separations, the depletion region do not overlap. (e) Two tip I-V curve for a tip separation of 50nm (red). At this distance, the MoS$_2$ surface shows a surface gap of 1.4eV. This curve can be fitted (black curve) by considering equation (2). At this probe separation, the depletion regions overlap resulting in the flat bands of MoS$_2$ as shown in (f).

Fig (2)

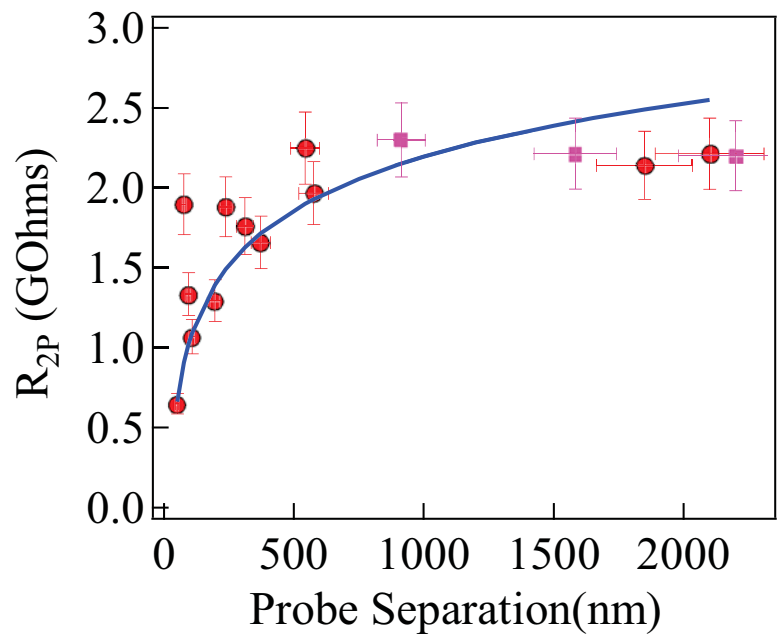

Fig (3) : Dependence of the series resistance ($R_{2P}$) with probe separation measured using two probes. Solid circles (red and pink) show the experimentally measured $R_{2P}$. The blue curve is the fit according to equation (3) assuming both tips with r = 30 nm and $R_C$ = 1.45 GΩ for the fit.

Fig (3)